\documentstyle[aps]{revtex}
\begin{document}
\newcommand{\beq}{\begin{equation}}
\newcommand{\eeq}{\end{equation}}
\newcommand{\bea}{\begin{eqnarray}}
\newcommand{\eea}{\end{eqnarray}}
\def\plumin{\underline{+}}
\def\minplu{{{\stackrel{\underline{\ \ }}{+}}}}
\def\bfr{{\bf r}}
\def\bfw{{\bf w}}
\def\bfc{{\bf c}}
\def\hpsi{\hat \psi (\bfr)}
\def\hpsid{\hat \psi^\dagger (\bfr)}
\def\tpsi{\tilde \psi (\bfr)}
\def\tpsid{\tilde \psi^\dagger (\bfr)}

\twocolumn[\hsize\textwidth\columnwidth\hsize\csname@twocolumnfalse%
\endcsname
\draft
%\begin{title}
%\bf
\title{Finite temperature excitations of a trapped Bose gas}
%\end{title}
\author{D. A. W. Hutchinson and E. Zaremba}
%\begin{instit}
\address{
Department of Physics,
Queen's University,
Kingston, Ontario, Canada K7L 3N6
}
\author{A. Griffin}
\address{
Department of Physics,
University of Toronto,
Toronto, Ontario, Canada M5S 1A7
}
%\end{instit}

\date{\today}

\maketitle

\begin{abstract}
We present a detailed study of the temperature
dependence of the condensate and noncondensate density profiles
of a Bose-condensed gas in a parabolic trap. These quantitites are
calculated self-consistently using the 
Hartree-Fock-Bogoliubov equations within the Popov approximation.
Below the Bose-Einstein transition the
excitation frequencies have a realtively weak temperature
dependence even though the condensate is strongly depleted.
As the condensate density goes to zero through the
transition, the excitation frequencies are strongly affected 
and approach the frequencies of a noninteracting gas in the high
temperature limit.
\end{abstract}

]

Since the recent discovery of Bose-Einstein condensation (BEC) in
ultracold trapped atomic gases\cite{anderson,davies}, 
there have been a number of calculations of
the collective excitations in such systems.
These calculations\cite{fetter72,edwards,singh,S96}
have mainly concentrated on the
excitations at $T =0$ using the standard Bogoliubov equations,
which assume that all the atoms are in the Bose 
condensate.  Recent experimental 
results for rubidium\cite{jin} and sodium\cite{mewes} are in
good agreement with these $T = 0$ calculations, which give the
frequencies as a function of the number of atoms. However, at 
finite temperatures where the condensate is strongly depleted, a
generalization of these theories is needed. The simplest theory
which includes the effect of the noncondensate atoms in a
self-consistent manner is the
Hartree-Fock-Bogoliubov (HFB) approximation\cite{griffin}.

In this Letter, we use a simplified version of the HFB
approximation originally introduced by Popov\cite{popov}
in the study of
uniform Bose gases close to the transition and discussed more
recently\cite{shi,griffin}
as a useful collisionless approximation over a wide
range of temperatures. The Popov approximation ignores the
anomalous (or off-diagonal) correlations present in the full
HFB theory. Within this approximation, we calculate {\it 
self-consistently}
the condensate $n_c(\bfr)$ and noncondensate $\tilde
n(\bfr)$ density profiles at $T = 0$ as well as at finite
temperatures. Our present results are the first self-consistent
calculation of these quantities within the HFB-Popov theory
which avoids the semiclassical approximation for the
quasiparticle excitations used previously\cite{giorgini}, or
various other simplifications\cite{goldman}. Our calculations
also provide the HFB-Popov excitation frequencies, which at low
temperatures are the collective modes of the condensate. We find
that these modes are weakly temperature-dependent in a range of
temperatures below $T_c$, even when the condensate is strongly
depleted. At temperatures above $T_c$, the lowest excitation 
frequencies approach the excitation energies of noninteracting
particles in a parabolic well.

Starting with the equation of motion for the Bose field operator
$\hpsi$, and its decomposition $\hpsi \equiv
\Phi(\bfr) + \tpsi$ in terms of condensate and
noncondensate parts\cite{fetter72,griffin}, 
the condensate wavefunction $\Phi(\bfr)$
is defined within the Popov approximation by the 
generalized Gross-Pitaevskii equation\cite{griffin}
\beq
\left [-{\nabla^2\over 2m} + V_{ext}(\bfr) + g[n_c(\bfr) +
2\tilde
n(\bfr) ] \right ]
\Phi(\bfr) = \mu \Phi(\bfr) \,.
\label{cond2}
\eeq
Here, $n_c(\bfr) \equiv |\Phi(\bfr)|^2$ and $\tilde
n(\bfr) \equiv \langle \tpsid \tpsi \rangle$ 
are the condensate and noncondensate densities, respectively.
The Popov approximation\cite{griffin,popov,shi}
consists of omitting the anomolous 
correlation $\langle \tpsi \tpsi \rangle$, but keeping $\tilde
n(\bfr)$. The condensate wavefunction in Eq.(\ref{cond2}) 
is normalized to
$N_c$, the total number of particles in the condensate.
$V_{ext}(\bfr)$ is the external confining potential and $g = 4\pi
\hbar^2a/m$ is the interaction strength
determined by the $s$-wave scattering length $a$. The 
condensate eigenvalue is given by
the chemical potential $\mu$\cite{comment}.

The usual Bogoliubov transformation, $\tpsi = \sum_i
[u_i(\bfr)\hat \alpha_i - v_i^*(\bfr)\hat \alpha_i^\dagger ]$,
to the new Bose operators $\hat \alpha_i$ and $\hat \alpha_i^\dagger$
leads to the  coupled HFB-Popov equations\cite{griffin}
\bea
\hat {\cal L} u_i(\bfr) - g n_c(\bfr) v_i(\bfr) &=& E_i
u_i(\bfr)
\nonumber \\
\hat {\cal L} v_i(\bfr) - g n_c(\bfr) u_i(\bfr) &=& -E_i
v_i(\bfr)\,,
\label{HFB}
\eea
with $\hat {\cal L} 
\equiv -\nabla^2/2m + V_{ext}(\bfr)+ 2gn(\bfr) - \mu
\equiv \hat h_0 + gn_c(\bfr)$. These
equations define the quasiparticle excitation energies $E_i$ and
the quasiparticle amplitudes $u_i$ and $v_i$. 
Once these quantities have been determined, the 
noncondensate density is obtained from the
expression\cite{fetter72,griffin}
\bea
\tilde n(\bfr) &=& \sum_i \left \{ |v_i(\bfr)|^2 +
\left [ |u_i(\bfr)|^2+|v_i(\bfr)|^2 \right ] N_0(E_i)
\right \}
\nonumber 
\\ &\equiv& \tilde n_1(\bfr)+\tilde n_2(\bfr)\,,
\label{tilden}
\eea
where $\tilde n_1(\bfr)$ is that part of the density which 
reduces to the ground-state noncondensate density as
$T \to 0$.  The component $\tilde n_2(\bfr)$ depends
upon the Bose distribution $N_0(E_i)=(e^{\beta E_i}-1)^{-1}$ and
vanishes in the $T \to 0$ limit.

Rather than solving the coupled equations in Eq.(\ref{HFB})
directly, we introduce a new method based on the
auxiliary functions
$\psi_i^{(\pm)}(\bfr) \equiv u_i(\bfr) \pm v_i(\bfr)$.
From Eq.(\ref{HFB}), these are found to be
solutions of the uncoupled equations
\bea
\hat h_0^2 \psi_i^{(+)}(\bfr) +2gn_c(\bfr) \hat h_0
\psi_i^{(+)}(\bfr) &=& E_i^2 \psi_i^{(+)}(\bfr) \label{psi+}
\label{8}\\
\hat h_0^2 \psi_i^{(-)}(\bfr) + 2g\hat h_0 n_c(\bfr)
\psi_i^{(-)}(\bfr) &=& E_i^2 \psi_i^{(-)}(\bfr)\,.
\label{9}
\eea
The two functions are related to each other by 
$\hat h_0 \psi_i^{(+)} = E_i \psi_i^{(-)}$.
We note that the collective modes of the condensate 
can be shown to have an
associated density fluctuation given by $\delta n_i(\bfr) 
\propto \Phi(\bfr) \psi_i^{(-)}(\bfr)$.

To solve Eq.(\ref{8}) we introduce the
normalized eigenfunction basis defined as the solutions of
$\hat h_0 \phi_\alpha(\bfr) = \varepsilon_\alpha
\phi_\alpha(\bfr)$. 
The lowest energy solution gives the condensate wavefunction 
$\Phi(\bfr) = \sqrt{N_0}\phi_0(\bfr)$ with eigenvalue
$\varepsilon_0 = 0$. 
Using the expansion $\psi_i^{(+)}(\bfr) = \sum_\alpha 
c_\alpha^{(i)} \phi_\alpha(\bfr)$ in Eq.(\ref{8}), we obtain the
eigenvalue equation\cite{zaremba}
\beq
\sum_\beta \left \{ M_{\alpha \beta} + \varepsilon_\alpha
\delta_{\alpha \beta} \right \} \varepsilon_\beta c_\beta^{(i)}
= E_i^2 c_\alpha^{(i)}\,,
\label{12}
\eeq
where the matrix $M_{\alpha
\beta} \equiv 2g\int d\bfr \, \phi_\alpha^*(\bfr)
n_c(\bfr)\phi_\beta(\bfr)$.
The orthonormality of the $u$ and $v$ functions\cite{fetter72}
in Eq.(\ref{HFB})
requires that the expansion coefficients $c_\alpha^{(i)}$ be 
normalised as
$\sum_\alpha \varepsilon_\alpha c_\alpha^{(i)*} c_\alpha^{(j)}
= E_i \delta_{ij}$.
The two parts of the noncondensate density $\tilde n(\bfr)$ as
defined in Eq.(\ref{tilden}) are given in
terms of the $c_\alpha^{(i)}$ through the expressions
\beq
\tilde n_1(\bfr) = {1\over 4} \sum_i \left |
\sum_\alpha \left
(1 - {\varepsilon_\alpha \over E_i} \right ) c_\alpha^{(i)}
\phi_\alpha(\bfr) \right |^2
\label{22}
\eeq
and
\bea
\tilde n_2(\bfr) &=& {1\over 2} \sum_i \left \{ \left \vert
\sum_\alpha
c_\alpha^{(i)} \phi_\alpha(\bfr) \right \vert^2
+ \left \vert
\sum_\alpha {\varepsilon_\alpha \over E_i} c_\alpha^{(i)}
\phi_\alpha(\bfr) \right \vert^2 \right \} \nonumber \\
& &\qquad \times N_0(E_i)\,.
\label{23}
\eea

The calculational procedure can be
summarized for an arbitrary confining potential as follows:
Eq.(\ref{cond2}) is first solved self-consistently for
$\Phi(\bfr)$, with $\tilde n(\bfr)$ set to zero.
Once $\Phi(\bfr)$ is known, the eigenfunctions of $\hat h_0$
required in the expansion of the excited state amplitudes are
generated numerically.  The matrix problem in Eq.(\ref{12}) 
is then set up to obtain the eigenvalues $E_i$, and the corresponding 
eigenvectors $c_\alpha^{(i)}$ are used to evaluate the 
noncondensate density.
This result is inserted into Eq.(\ref{cond2}) and the
process is iterated to convergence. At each step, the
normalization of the condensate wave function is
defined by $\int d\bfr |\Phi(\bfr)|^2 = N - \tilde N$, where 
$\tilde N$ is the total number of noncondensate particles.

As an illustration of this procedure, we consider an
isotropic harmonic potential $V_{ext}(\bfr) = {1\over 2}m\omega_0^2 r^2
$, for which the order parameter
$\Phi(\bfr)$ is a spherically symmetric function. In addition,
we use the following parameters\cite{anderson,jin}: 
$m(^{87}Rb) = 1.44 \times 10^{-25}$ kg,
$\nu_0 = \omega_0/2\pi = 200$ Hz and an $s$-wave  scattering
length of $a \simeq 110 a_0 = 5.82 \times 10^{-9}$ m. 
Throughout we express lengths and
energies in terms of the characteristic oscillator length 
$d = (\hbar/m\omega_0)^{1/2} = 7.62 \times 10^{-7}$ m 
and the characteristic trap energy $\hbar \omega_0
= 1.32 \times 10^{-31}$ J, respectively. A convenient
dimensionless parameter\cite{edwards,singh,S96} describing the
effective strength of the interactions is $\gamma \equiv Na/d$,
which is proportional to the ratio of the average interaction
energy $g\bar n$ (where $\bar n = N/d^3$) to the characteristic
energy level spacing $\hbar \omega_0$. In the absence of
$\tilde n(\bfr)$ in Eqs.(\ref{cond2}) and (\ref{HFB}), the
various properties of the condensate scale\cite{edwards,singh}
with this single parameter $\gamma$.

For a spherical trap, the eigenfunctions of $\hat h_0$ and the
quasiparticle excitations can be classified according to the
angular momentum $l$. The number of excitations generated for a
given $l$ is determined by the dimension of the eigenfunction
expansion and is chosen sufficiently large to
ensure convergence of the noncondensate density as determined by
the sum over all modes in Eqs.(\ref{22}) and (\ref{23}).
The noncondensate 
density profile is shown as a sum of contributions from each
angular momentum $l$ in Fig. 1(a) for $N = 2000$ atoms. The
central core of the noncondensate is made up from the set of
excitations with $l = 0$.

The fraction of atoms in the noncondensate 
at $T = 0$ is shown in Fig. 1(b)
as a function of the number of atoms $N$ in the droplet. This
fraction does not exceed 0.5\% of the total at $T = 0$ for the
range of $N$ considered and the parameters given above. 
Having most of the atoms
in the condensate is in sharp contrast with the
situation in superfluid He$^4$ where 90\%
of the atoms are outside the condensate at $T = 0$.
The monotonic increase of $\tilde N$ with $N$ is due to the
dependence of the coupling between the $u$ and $v$ functions in 
the HFB-Popov equations on the effective interaction strength
$\gamma$. The latter increases with $N$ and therefore leads to
an increasing number of excitations out of the condensate.
Since $\tilde N$ is a small fraction of the total at 
$T = 0$, it is to a very good approximation a function
of the single parameter $\gamma$. The value of $\tilde N$ for
other interaction strengths and harmonic potentials can therefore
be obtained from the results in Fig. 1(b) by a simple scaling
with $\gamma$.

With increasing temperature, the noncondensate density increases
by virtue of the Bose distribution term in Eq.(\ref{tilden}). 
We note that $\tilde n_1(\bfr)$ depends weakly on temperature 
and is rapidly dominated by $\tilde n_2(\bfr)$ at elevated 
temperatures. In Fig. 2(a), we show our self-consistent
results for $\tilde n(\bfr)$ for $N = 2000$ atoms for a range of
temperatures below the transition temperature of approximately
100 nK (see below). The insert to Fig. 2 compares $\tilde
n(\bfr)$ and $n_c(\bfr)$ at $T = 75nK$.
The trapped atomic gas has a two-component 
structure\cite{anderson,goldman} at elevated temperatures,
with a dense core of condensed atoms sitting on top
of a diffuse gaseous cloud of excited atoms with a long 
tail\cite{comment2}.
Although the noncondensate density in Fig.2(b) looks small in 
comparison to the condensate, both have an approximately
equal number of atoms at this temperature because of the $r^2$ 
weighting of the integrated density.
We also see that $\tilde n(\bfr)$ develops a peak at the edge of
the condensate with increasing temperature. This is the analogue
of the sharp peak found in the semi-classical Thomas-Fermi 
approximation\cite{goldman} in which the
condensate has a sharp cutoff at some radius $R_0$.

In Fig. 3(a) we show the total number of atoms $N_c$ in the 
condensate as a function of temperature for a 2000 atom droplet.
It can be seen that $N_c$ is falling to zero at
approximately 100 nK, which is close to the BEC transition
temperature $T_c$ predicted by the semiclassical
approximation\cite{Groot,bagnato}. Although there is no sharp
transition for a finite system, there is still a characteristic
temperature above which $N_c$ is small. One way of defining
this temperature is to fit the temperature variation of $N_c$
to the functional form
$N_c(T) = N_c(0)[1-(T/T_c)^\alpha]$, treating $T_c$ and $\alpha$ as
fitting parameters. This functional
form provides a reasonable parameterization over a wide
intermediate range of temperatures below $T_c$. The values of
$T_c$ extracted in this way are found to scale approximately as
$N^{1/3}$, which is the $N$-dependence 
found for a trapped, noninteracting Bose gas. 
As can be seen from Fig. 3(b), the
numerical values of $T_c$ found here are 
slightly smaller than the ideal gas values, in agreement 
with recent theoretical\cite{giorgini} and
experimental\cite{mewes2} results. The value of
$\alpha$ extracted from the fits is 2.3, as compared to the ideal
Bose gas value\cite{bagnato} of 3.

The size of the noncondensate $\tilde n(\bfr)$ is, of course,
determined by the quasiparticle excitations which deplete the
condensate. The mode frequencies 
for the lowest modes of angular momentum $l = 0$, 1 and 2 are 
shown in Fig. 4(a) for $T = 0$. The $l = 0$ and  $l = 2$ 
mode frequencies split apart from the degenerate harmonic
potential eigenvalue as the effective interaction strength
$\gamma$ (or $N$, for fixed $a/d$) is increased.
The $l = 0$ mode is a ``breathing-type" mode which is influenced
by the
compressibility of the condensate, and increases in frequency as
a result of the repulsive interactions between the atoms. On the
other hand, the $l = 2$ quadrupole mode is a shape resonance
controlled by the surface tension of the drop and is seen to 
decrease in frequency with increasing $N$. Due to the small
magnitude of the noncondensate, these frequencies are very similar 
to the recent results at $T =0$ of Singh and Rokhsar\cite{singh},
Stringari\cite{S96} and others\cite{edwards} in which the
noncondensate was ignored. 

The lowest lying $l = 1$ mode is the center-of-mass 
mode of the condensate. It lies very close to, but not exactly
at, $\omega_0$.
According to the generalized Kohn theorem for parabolic
confinement\cite{GKT}, one would expect a $l = 1$ mode at
precisely
$\omega_0$ corresponding to a rigid oscillation of the total droplet 
density. This property is satisfied in the Bogoliubov 
approximation\cite{singh,S96} in which all particles are in the
condensate. However, in the HFB-Popov approximation, the 
condensate is effectively moving in the presence of the
`external' potential $V_{ext}(\bfr)+2g\tilde n(\bfr)$, which
deviates from the ideal parabolic form. It is only because
$2g\tilde n(\bfr)$ is a small perturbation in the present case
that the generalized Kohn theorem is so nearly satisfied.
An improved approximation in which the dynamics of the 
noncondensate is treated on an equal footing with that of the 
condensate is needed to recover the true
center-of-mass mode in which both components of the density
oscillate together.

We have also examined the quasiparticle excitations as a function 
of temperature. Due to convergence difficulties at the higher
temperatures\cite{comment2}, we show in Fig. 4(b) results for a 
smaller droplet containing $N = 200$ atoms, which has a 
transition in the middle of the accessible temperature range.
The behaviour for larger drops is qualitatively 
similar\cite{comment3}. For temperatures below
the BEC transition, the Bose broken symmetry leads quite generally 
to the equivalence of the single-particle and density fluctuation
spectra\cite{griffin/book}. However, in the normal phase,
this equivalence no longer holds and the HFB single-particle 
excitations shown in Fig.4(b) above $T_c$ cannot be identified with
the density fluctuation spectrum. We observe that the splitting
between the $l = 0$ and $l = 2$ modes decreases as $n_c(\bfr)$
goes to zero through the transition temperature, and both modes
tend to the harmonic excitation frequency of 2$\omega_0$. The $l
= 1$ dipole excitation remains close to $\omega_0$ for all 
temperatures. The explanation of the high temperature behaviour
is straightforward.
Once $n_c(\bfr)$ goes to zero, the single-particle excitations 
reduce to the normal HF excitations for the self-consistent
potential $V_{ext}(\bfr) + 2gn(\bfr)$. Since the total density 
$n(\bfr)$ becomes spread out at temperatures above $T_c$,
the HF interaction potential provides a fairly constant potential 
energy shift to the harmonic confining potential and the lowest
excitation energies reduce to those of a noninteracting gas in
a harmonic well.

In conclusion, we have presented a detailed study of the
behaviour of the condensate and noncondensate density profiles
as a function of temperature within the Popov approximation to the
full HFB equations of motion\cite{griffin}. Although we have
only considered isotropic parabolic wells, our method based on
Eqs.(\ref{8})-(\ref{12}) can also be used for anisotropic, 
as well as anharmonic, traps. Previous theoretical
studies\cite{goldman,chou,giorgini} have made use of
simplified semiclassical approximations, whereas our results are
based on a fully self-consistent numerical treatment of the
HFB-Popov equations. We have also determined the temperature
dependence of the excitations through the BEC transition and
find an interesting evolution from collective to 
single-particle-like behaviour. Our work
emphasizes the need for more experimental studies in this
temperature range. A study of the full HFB equations
including the off-diagonal correlations neglected in the Popov
approximation will be reported elsewhere.

This work was supported by grants from the Natural Sciences and
Engineering Research Council of Canada.

\begin{figure}
\caption{
(a) Noncondensate density at $T=0$ for 2000 rubidium atoms in an
isotropic parabolic trap.
The innermost curve corresponds to the $l =0$
excitations, and each successive curve includes the contribution
of the next angular momentum.
The insert (b) shows the variation of the 
noncondensate fraction (in percent) at $T=0$ as a function of $N$.
}
\end{figure}

\begin{figure}
\caption{
(a) Noncondensate density for 2000 rubidium
atoms at various temperatures.
(b) The noncondensate (solid) and
condensate (dashed) densities at $T=75$ nK.
}
\end{figure}

\begin{figure}
\caption{
(a) The number of atoms within the condensate for $N=2000$,
as a function of temperature. (b) The critical temperature vs $N$
as calculated self-consistently using the HFB-Popov equations 
(solid line) compared to the ideal Bose gas result
$T_c = (\hbar \omega_0 /k_B) (N/\zeta (3))^{1/3}$ (dashed line).
}
\end{figure}

\begin{figure}
\caption{
(a) Lowest mode frequencies (in units of the trap frequency 
$\omega_0$) for
angular momenta $l=0$, 1 and 2, as a function of the number of
atoms in the trap at $T=0$. (b) Temperature variation of the
frequencies for $N= 200$. The lower part of the panel shows
$N_c$ vs $T$.
}
\end{figure}


\begin{references}
\bibitem{anderson} M. H. Anderson et al., Science {\bf 269}, 
198 (1995).
\bibitem{davies} K. B. Davies et al., Phys. Rev. Lett. {\bf 75},
3969 (1995).
\bibitem{fetter72} A. L. Fetter, Ann. Phys. (N.Y.) {\bf 70}, 67
(1972).
\bibitem{edwards} M. Edwards et al.,
Phys. Rev. Lett. {\bf 77}, 1671 (1996).
\bibitem{singh} K. G. Singh and D. S. Rokhsar, Phys. Rev. Lett.
{\bf 77}, 1667 (1996).
\bibitem{S96} S. Stringari, Phys. Rev. Lett. {\bf 77}, 2360
(1996).
\bibitem{jin} D. S. Jin et al.,
Phys. Rev. Lett. {\bf 77}, 420 (1996).
\bibitem{mewes} M.-O. Mewes et al.,
Phys. Rev. Lett. {\bf 77}, 992 (1996).
\bibitem{griffin} A. Griffin, Phys. Rev. B {\bf 53}, 9341
(1996) and references therein.
\bibitem{popov} V. N. Popov, {\it Functional Integrals and
Collective Modes} (Cambridge University Press, New York, 1987),
Ch. 6.
\bibitem{shi} H. Shi, G. Verachaka and A. Griffin, Phys. Rev.
B {\bf 50}, 1119 (1994).
\bibitem{giorgini} S. Giorgini, L. P. Pitaevskii and S.
Stringari, preprint (cond-mat/9607117); and private communication.
\bibitem{goldman} V. V. Goldman, I. F. Silvera and A. J.
Leggett, Phys. Rev. B {\bf 24}, 2870 (1981).
\bibitem{comment} This is strictly true only when the number of
noncondensate particles is a small fraction of the total. Near
and above $T_c$, the chemical potential is fixed as in
Ref.\cite{goldman}.
\bibitem{zaremba} An analogous formalism has been developed in
the study of electronic magnetoplasmons: E. Zaremba and H.
C. Tso, Phys. Rev. B {\bf 49}, 8147 (1994) and E. Zaremba, Phys.
Rev. B {\bf 53}, 10512 (1996).
\bibitem{comment2}
The tail is made up of higher angular momentum states and its
accurate calculation is limited at higher temperatures by the
necessary truncation of the numerical basis set.
For $N = 2000$, the error in the total noncondensate fraction
begins to exceed a few percent at $0.8T_c$.
\bibitem{Groot} S. R. de Groot, G. J. Hooyman and C. A. Ten
Seldam, Proc. R. Soc. (London), series A {\bf 203}, 266 (1950).
\bibitem{bagnato} V. Bagnato, D. E. Pritchard and D. Kleppner,
Phys. Rev. A {\bf 35}, 4354 (1987).
\bibitem{mewes2} M.-O. Mewes et al., Phys. Rev. Lett. 
{\bf 77}, 416 (1996).
\bibitem{GKT} See J. F.  Dobson, Phys. Rev. Lett. {\bf 73}, 
2244 (1994) and references therein.
\bibitem{griffin/book} A. Griffin, {\it Excitations in a
Bose-Condensed Liquid} (Cambridge, New York, 1993), Ch. 5 and
references therein.
\bibitem{chou} T. T. Chou, C. N. Yang and L. H. Yu, Phys. Rev. A
{\bf 53}, 4257 (1996).
\bibitem{comment3}
Variations in the quasiparticle excitation energies in 
Eq.(\ref{HFB}) are mainly determined by variations in 
$n_c(\bfr)$, which depends on both $N$ and $T$. The weaker
$N$-dependence of the frequencies with increasing $N$ (see
Fig.4(a)) implies a similar weakening of the temperature
dependence below $T_c$ for larger $N$.
\end{references}
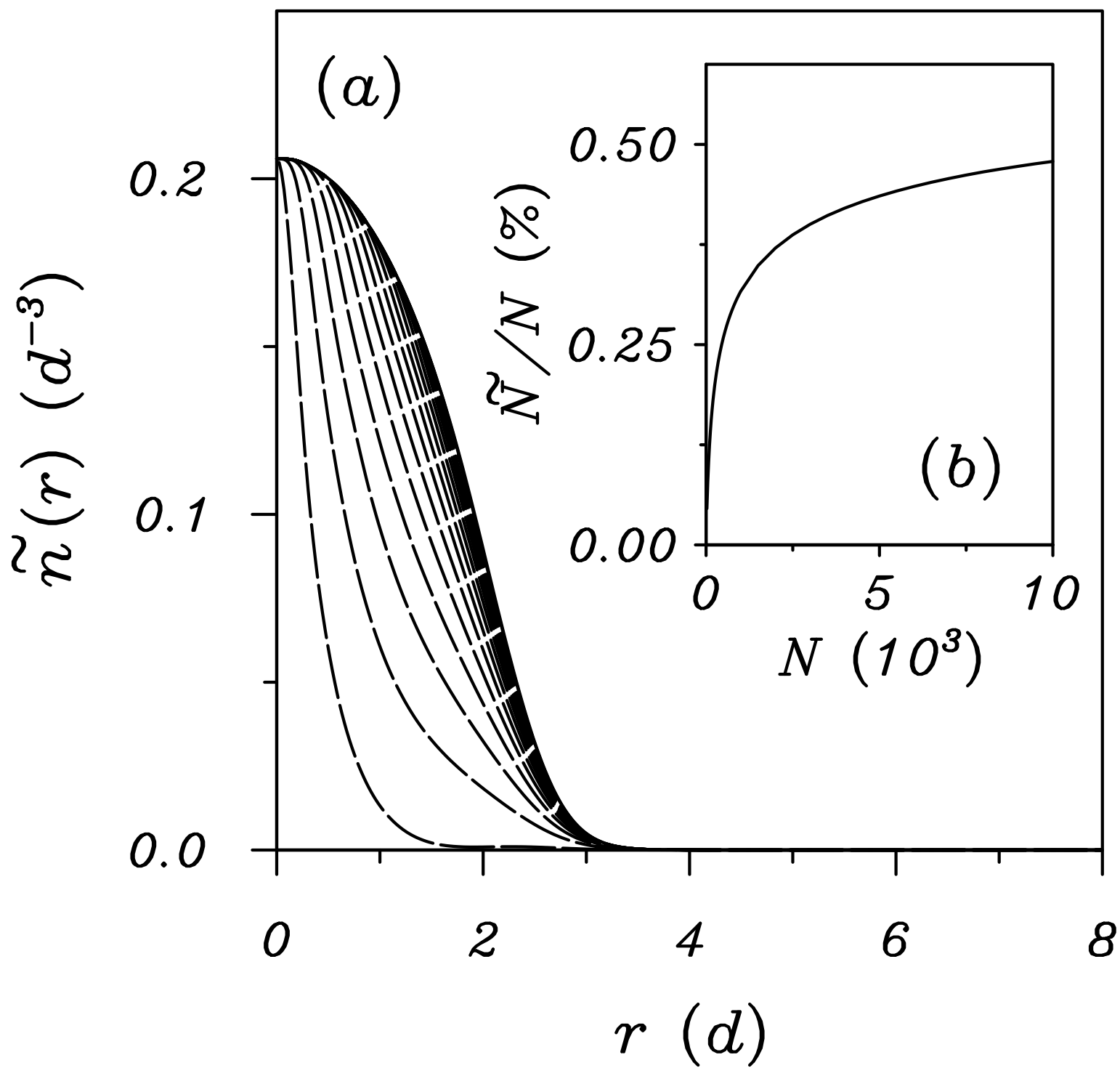
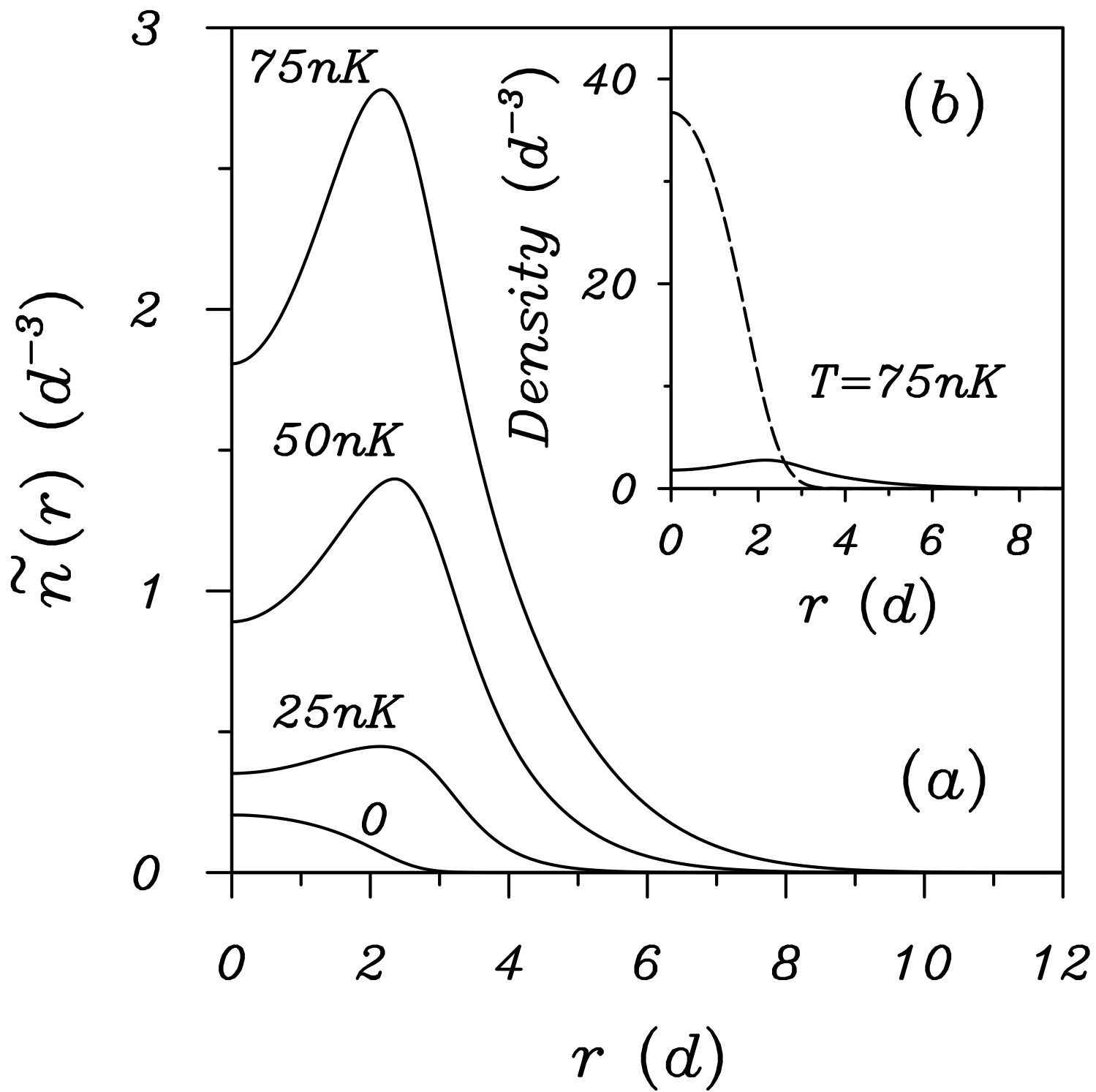
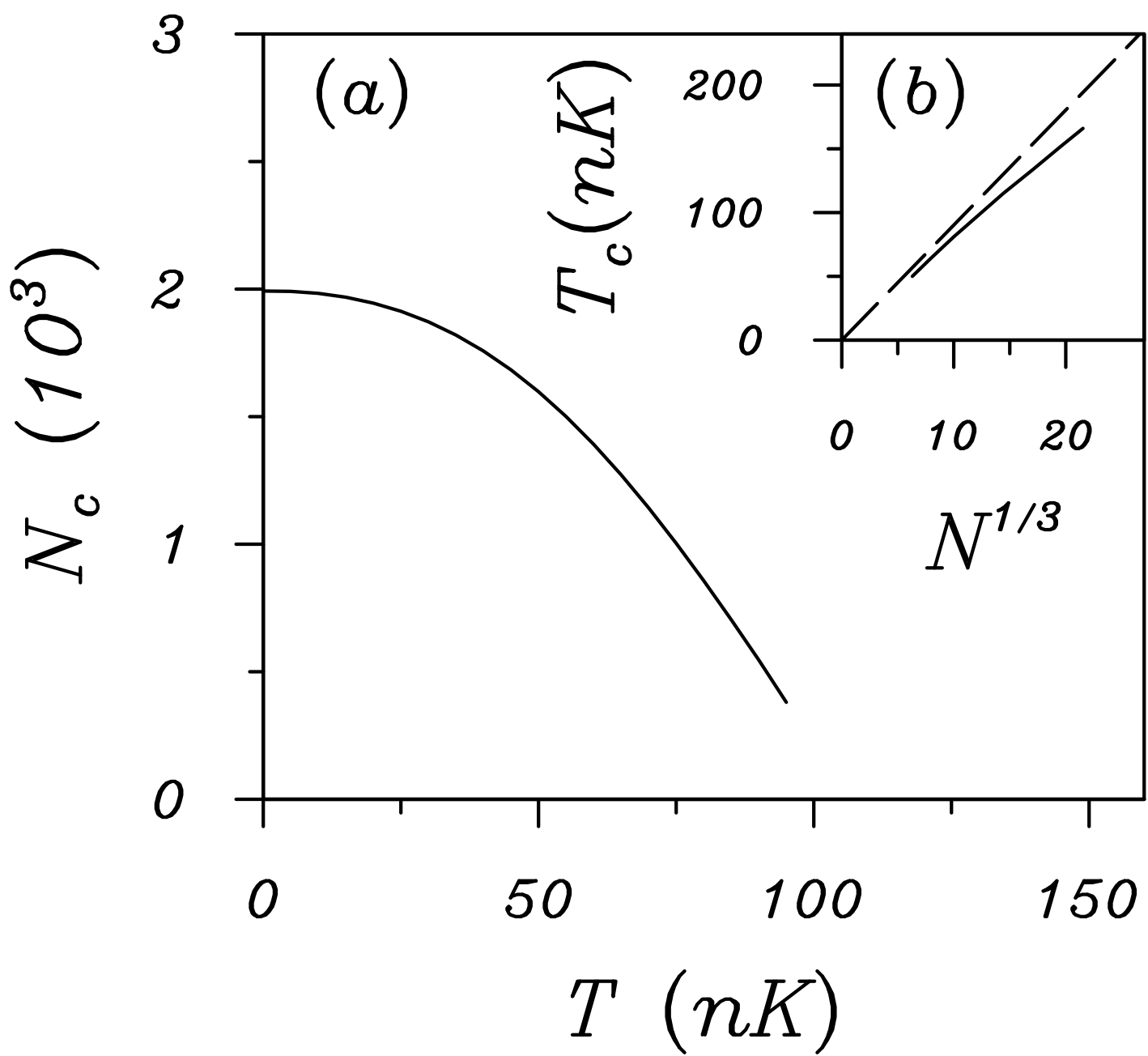
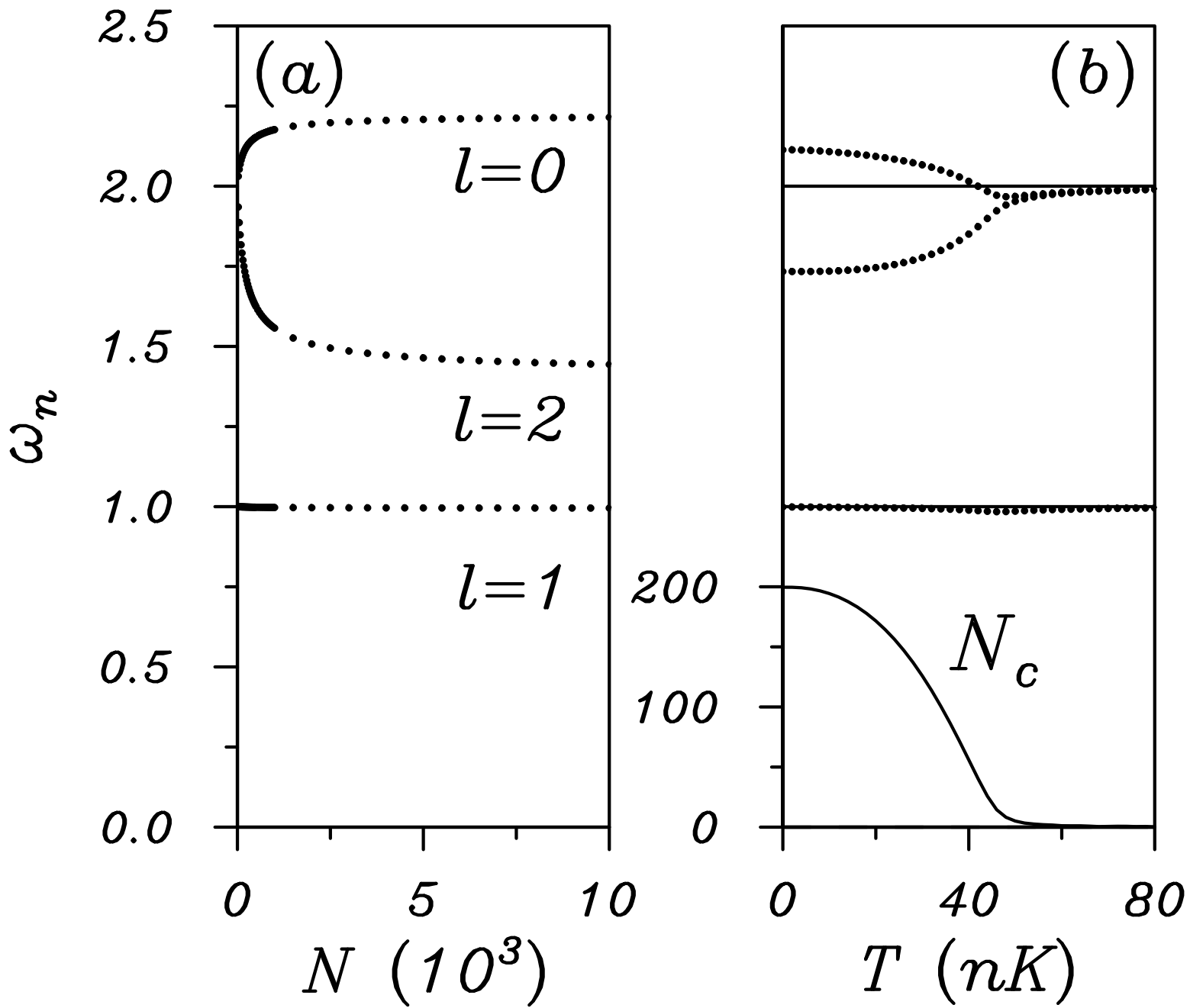
\end{document}